\documentclass[iop,apjl]{emulateapj}
\usepackage{txfonts}
\usepackage{microtype}

\usepackage{graphicx}

\usepackage{snapshot}

\usepackage{color}
\definecolor{hcolor}{rgb}{0.,0.,0.8}
\usepackage[colorlinks=true,citecolor=hcolor,filecolor=hcolor,
linkcolor=hcolor, urlcolor=hcolor,
pdftitle={Sco~X-1 VHE observations with the MAGIC telescopes},
pdfauthor={The MAGIC Collaboration}]{hyperref}

\shorttitle{Scorpius~X-1 VHE observations with the MAGIC telescopes}
\shortauthors{Aleksi\'c et al.}

\journalinfo{The Astrophysical Journal Letters, 735:L5 (5pp), 2011 July
1\hfill
doi:10.1088/2041-8205/735/1/L5}
\submitted{Submitted 2011 March 29; accepted 2011 May 17; published 2011
June 3}

\newcommand{\cms}{\ensuremath{\mathrm{cm^{-2}\,s^{-1}}}}
\newcommand{\ergcms}{\ensuremath{\mathrm{erg\,cm^{-2}\,s^{-1}}}}
 
\begin{document}

\title{A search for Very High Energy gamma-ray emission from Scorpius
X-1\\ with the MAGIC telescopes}

%
\author{
J.~Aleksi\'c$^{1}$,
E.~A.~Alvarez$^{2}$,
L.~A.~Antonelli$^{3}$,
P.~Antoranz$^{4}$,
M.~Asensio$^{2}$,
M.~Backes$^{5}$,
J.~A.~Barrio$^{2}$,
D.~Bastieri$^{6}$,
J.~Becerra~Gonz\'alez$^{7,8}$,
W.~Bednarek$^{9}$
% ...
}
\affil{$^{1}$ IFAE, Edifici Cn., Campus UAB, E-08193 Bellaterra, Spain}
\affil{$^{2}$ Universidad Complutense, E-28040 Madrid, Spain}
\affil{$^{3}$ INAF National Institute for Astrophysics, I-00136 Rome, Italy}
\affil{$^{4}$ Universit\`a  di Siena, and INFN Pisa, I-53100 Siena, Italy}
\affil{$^{5}$ Technische Universit\"at Dortmund, D-44221 Dortmund, Germany}
\affil{$^{6}$ Universit\`a di Padova and INFN, I-35131 Padova, Italy}
\affil{$^{7}$ Inst. de Astrof\'{\i}sica de Canarias, E-38200 La Laguna, Tenerife, Spain}
\affil{$^{8}$ Depto. de Astrof\'{\i}sica, Universidad de La Laguna, E-38206 La Laguna, Spain}
\affil{$^{9}$ University of \L\'od\'z, PL-90236 Lodz, Poland}
% ...

\begin{abstract}
    The acceleration of particles up to GeV or higher energies in microquasars
    has been the subject of considerable theoretical and observational efforts
    in the past few years. Sco~X-1 is a microquasar from which evidence of
    highly energetic particles in the jet has been found when it is in the
    so-called Horizontal Branch (HB), a state when the radio and hard X-ray
    fluxes are higher and a powerful relativistic jet is present. Here we
    present the first very high energy gamma-ray observations of Sco~X-1,
    obtained with the MAGIC telescopes. An analysis of the whole dataset does
    not yield a significant signal, with 95\% CL flux upper limits above
    300\,GeV at the level of $2.4\times10^{-12}\,\cms$. Simultaneous
    \emph{RXTE} observations were conducted to provide the X-ray state of the
    source. A selection of the gamma-ray data obtained during the HB
    based on the X-ray colors did not yield a signal either, with an
    upper limit of $3.4\times10^{-12}\,\cms$. These upper limits place a
    constraint on the maximum TeV luminosity to non-thermal X-ray luminosity of
    $L_\mathrm{VHE}/L_\mathrm{ntX}\lesssim0.02$, that can be related to a
    maximum TeV luminosity to jet power ratio of
    $L_\mathrm{VHE}/L_\mathrm{j}\lesssim10^{-3}$. Our upper limits indicate
    that the underlying high-energy emission physics in Sco~X-1 must be
    inherently different from that of the hitherto detected gamma-ray
    binaries.
\end{abstract}


\section{Introduction}

It has been proposed that particles accelerated in relativistic microquasar
ejections could produce detectable gamma-ray emission
\citep{1999MNRAS.302..253A}. A confirmation of such predictions may be found in
Cygnus~X-3, an accreting X-ray binary from which non-thermal emission up to
energies above 100~MeV has been clearly detected
\citep{2009Natur.462..620T,2009Sci...326.1512F}. However, an extensive
observational campaign in the very high energy (VHE) gamma-ray band yielded no
detection \citep{2010ApJ...721..843A}. Another well-known microquasar,
GRS~1915+105, was observed at VHE but only upper limits were obtained
\citep{2009arXiv0907.1017S,2009A&A...508.1135H}. Finally, MAGIC found evidence
(at a post-trial significance level of $4.1\sigma$) for VHE gamma-ray emission
from the microquasar Cygnus~X-1 \citep{2007ApJ...665L..51A}, but this has not
yet been confirmed through an independent detection.

On the other hand, there are three {binary systems} that have been
unambiguously detected at TeV energies: LS\,5039 \citep{2005Sci...309..746A},
LS\,I\,$+$61\,303 \citep{2006Sci...312.1771A}, and PSR\,B1259$-$63
\citep{2005A&A...442....1A}. However, none of them can be clearly classified as
a microquasar. The nature of the compact object in LS\,5039 and
LS\,I\,$+$61\,303 is unknown: although it was originally proposed that they
were accreting microquasars \citep{2000Sci...288.2340P, 2004A&A...414L...1M},
there is growing evidence that they might contain a young non-accreting pulsar
\citep{2006A&A...456..801D}, as is also the case for PSR\,B1259$-$63
\citep{1992ApJ...387L..37J}. A fourth candidate is HESS~J0632$+$057, a galactic
plane point-like TeV source with variable radio, X-ray and TeV emission
\citep[and references therein]{2009MNRAS.399..317S}. {Recent observations
have resulted in the detection of a $320\pm5$\,day X-ray period
\citep{2011arXiv1104.4519B}, and detection of slightly extended radio emission
at milliarcsecond scales \citep{2011ATel.3180....1M} similar to that found in
all other gamma-ray binaries.}

In the case of the microquasar Cygnus~X-1, the massive, luminous stellar companion
provides an intense target photon field for inverse Compton (IC) scattering.
However, this photon field could also absorb the produced gamma-rays through
pair production with opacities up to 10 at 1~TeV \citep{2007A&A...464..437B}.
Such strong absorption might not be present in the case of low-mass X-ray
binaries (LMXBs), where the companion star {has a lower mass and,
correspondingly,} a lower luminosity. 
Low-mass systems with persistent accretion and powerful
relativistic jets could produce TeV emission via synchrotron self-Compton (SSC),
and would be good candidates to be detected at these energies given their
continuous activity and the lack of severe absorption. 
However, external IC cannot be fully discarded in LMXBs given strong accretion
disk emission \citep{2006A&A...447..263B} or X-ray-enhanced radiation from the
stellar companion, which would also increase the TeV opacities
\citep{2010A&A...514A..61B}.

The Z sources are a class of LMXBs that contain a low magnetic field neutron star
accreting close to the Eddington limit. Their X-ray intensities and
colors change on timescales of hours, and their paths in a hard color versus
soft color diagram follow roughly Z-shaped tracks \citep{1989A&A...225...79H}.
Along this track the sources change between different spectral states known as
Horizontal Branch (HB), Normal Branch (NB), and Flaring Branch (FB). A
traditional interpretation of these different X-ray states is a variation in the
mass accretion rate. Recent evidence, however, points towards constant mass
accretion rate along the Z track \citep{2010ApJ...719..201H}, whereas different
mass accretion rates would give rise to the different LMXB subclasses. Since
most LMXBs have almost circular orbits, the X-ray states are not
expected to have an orbital dependence. Even though known as a Z track, with
the HB being the upper-left horizontal part of the track, in some sources
{the tracks may be more accurately described by} a double banana shape,
with the upper banana being the part corresponding to the HB/NB states and the
lower banana corresponding to the NB/FB states \citep{1989A&A...225...79H}.
Radio emission and hard X-ray power-law tails, strong evidence of particle
acceleration up to very high energies, have been only detected while the sources
are in the HB \citep{1990ApJ...365..681H,2006ApJ...649L..91D}. 

Sco~X-1, located at $2.8\pm0.3$\,kpc, is a prototype Z-type LMXB with an orbital
period of 0.787\,days that contains a low magnetic field neutron star and a
0.4\,M$_{\odot}$ M star. The circular binary orbit has an inclination of
44$^\circ$ and a separation of $\sim1.5\times10^{11}$\,cm (see
\citealt{2001ApJ...558..283F}, \citealt{2002ApJ...568..273S}, and references
therein).
Accretion takes place via Roche lobe overflow. The source displays a double
banana track in color-color diagrams (CD) and covers the whole track in a few
tens of hours, spending roughly half of the time in the HB. Spectral fits at hard
X-rays reveal a non-thermal power law with higher fluxes during the HB, reaching
values in the range 10$^{-10}$--10$^{-9}$\,\ergcms, and no high-energy cutoff
\citep{2006ApJ...649L..91D,2007ApJ...667..411D}. Moreover, twin relativistic
radio lobes moving at $\sim$0.5$c$ from the central source have been detected,
while successive flaring of the core and lobes reveals the action of an unseen,
highly relativistic flow with a speed above 0.95$c$ \citep{2001ApJ...558..283F}.
All these results clearly indicate the injection of highly energetic particles
when the source is in the HB, and suggest that the IC (likely SSC) process could
be at work when a powerful jet is present in this spectral state. The IC process
could generate VHE emission but, to adequately assess its origin, simultaneous
X-ray observations are required to monitor the source X-ray state. Sco~X-1
was claimed to be a source of TeV and PeV gamma-rays
\citep{1990A&A...232..383B,1991PhRvL..67.2248T}, but the low significance and
lack of later confirmation shed doubts on this evidence.

In 2010 May
we carried out a simultaneous observation campaign with the Cherenkov VHE
gamma-ray telescopes MAGIC and the X-ray observatory \emph{RXTE}. Here we
present the results of the first VHE observations of Sco~X-1 for selected X-ray
states, focusing on the HB state, where evidence for relativistic particles in
the jet has already been found.

\section{Observations and Data Analysis}

\subsection{MAGIC}

MAGIC consists of two 17~m diameter Imaging Atmospheric Cherenkov Telescopes
located at the Roque de los Muchachos Observatory on the Canary Island of La
Palma ($28^\circ$\,N, $18^\circ$\,W, 2200\,m). It became a stereoscopic system
in autumn 2009. Since then, the instrument sensitivity almost doubled with
respect to the one of the stand-alone telescope operation mode,
and currently it yields, at low zenith angles, 5$\sigma$ significance
detections above 250 GeV of fluxes as low as 0.8\% of the Crab Nebula
flux in 50\,hr \citep{2009arXiv0907.0960C,2011magic-inprep}.

{MAGIC observed Scorpius X-1 at high zenith angles, between 43$^\circ$ and
50$^\circ$, for a total amount of 7.75~hr during six consecutive nights
in 2010 May. Table~\ref{tab:log} shows the detailed observation log.}

\begin{deluxetable}{cccccc}
\tablecaption{Log of the VHE Observations.\label{tab:log}}
\tablehead{
\colhead{Date\tablenotemark{a}} & \colhead{Orbital} & \colhead{Eff.~Time} &
\colhead{X-ray State} &
\multicolumn{2}{c}{UL $(>300\,\mathrm{GeV})$\tablenotemark{b}} \\
\colhead{(MJD)} &  \colhead{Phase} & \colhead{(hr)} &                      &
\colhead{(\cms)} & \colhead{C.U.}
}
\startdata
55331.09 & 0.26--0.38 & 1.28 & NB/FB & $7.9\times10^{-12}$ & 6.4\%\\
55332.11 & 0.58--0.64 & 0.48 & HB    & $5.2\times10^{-12}$ & 4.2\%\\
55333.07 & 0.77--0.86 & 1.65 & NB/FB & $3.5\times10^{-12}$ & 2.8\%\\
55334.06 & 0.03--0.12 & 1.37 & HB    & $5.1\times10^{-12}$ & 4.1\%\\
55335.05 & 0.25--0.39 & 1.73 & HB    & $5.3\times10^{-12}$ & 4.3\%\\
55336.06 & 0.52--0.66 & 1.24 & NB    & $2.0\times10^{-12}$ & 1.6\%
\enddata
\tablenotetext{a}{The central MJD of each observation is quoted.}
\tablenotetext{b}{Upper limits at 95\% confidence level.\vspace{1em}}

\end{deluxetable}

{Data analysis was performed using the standard MAGIC analysis software. 
Each telescope records only the events selected by the hardware stereo trigger.
The obtained images are reconstructed, and parameterized
\citep{2009APh....30..293A}. The two images from the same stereo
event are combined, and the shower direction is determined as the intersection
of the corresponding single-telescope directions \citep{2010A&A...524A..77A}.
The background rejection relies on the definition of a global variable, called
\emph{hadronness}, which is computed by means of a Random Forest algorithm
\citep{2008NIMPA.588..424A}. The $\gamma$-ray signal is estimated through the
distribution of the squared angular distance between the reconstructed and the
catalog source position ($\theta^2$). The energy of each event is estimated by
using look-up tables created from Monte Carlo simulated $\gamma$-ray events.
The sensitivity of this high-zenith-angle analysis is 1.1\% of the Crab Nebula
flux for energies above 300 GeV in 50\,hr.}

For the case of non-detections, we computed flux upper limits using the method
of \cite{2005NIMPA.551..493R} including a 30\% systematic uncertainty. The
source spectrum was assumed to be a power law with a photon index of $\Gamma=3$.
This photon index was taken to account for the possibility that the
spectrum of Sco~X-1 is steeper than the Crab Nebula one, as would be the case
for some of the plausible emission scenarios (see Section~\ref{sec:disc}).
However, it must be noted that a 30\% change in the assumed photon index yields
a variation of less than 1\% in the flux upper limits. The integral upper
limits {apply to the photon fluxes from Sco~X-1 at energies above 300 GeV and
correspond to a confidence level of 95\%. A calculation of energy flux upper
limits from the differential photon flux upper limits yields a variation of
$\sim$1.5\% when considering a 30\% change in the assumed photon index.}

\subsection{X-rays}

\begin{figure}
    \centering
    \includegraphics{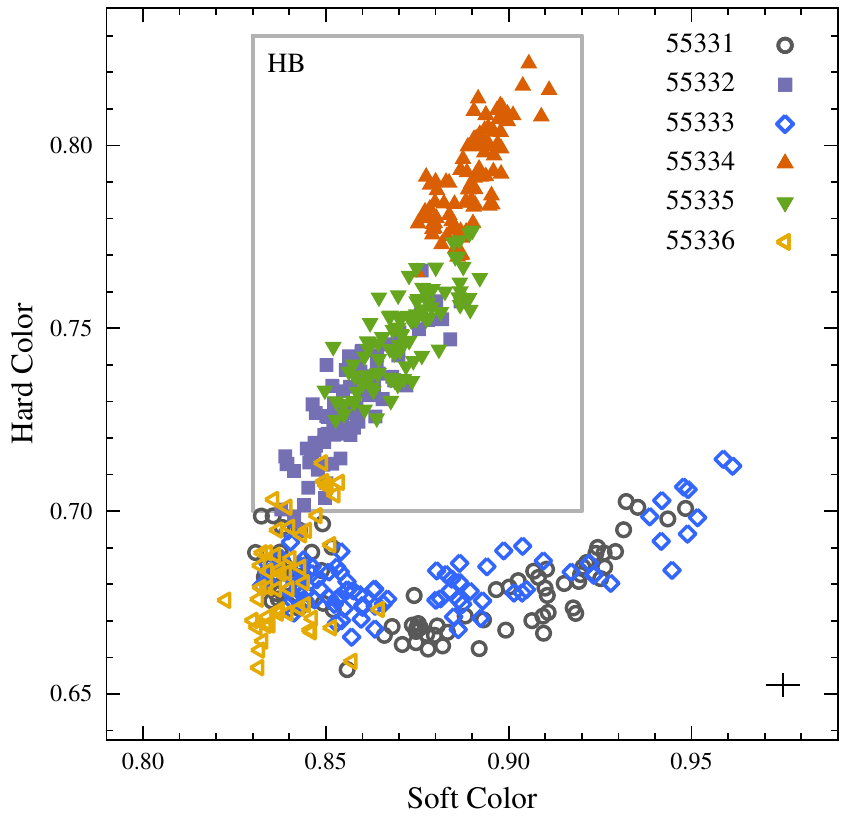}
    \caption{X-ray color-color diagram of Sco~X-1 from the \emph{RXTE}/PCA data
    obtained during the simultaneous MAGIC campaign. See the text for a
    definition of the Hard and Soft colors. The typical relative error of each
    measurement is around 0.5\% and is shown by a black cross at the
    bottom-right corner. The HB selection range is
    indicated by the gray box. The different symbols and colors indicate each of
    the different observation days, labeled in MJD. The filled symbols indicate
    days selected as HB and empty symbols days selected as NB or FB.}
    \label{fig:ztrack}
\end{figure}

Sco~X-1 was observed with \emph{RXTE} simultaneously {with} the MAGIC VHE
gamma-ray observations. To study the source X-ray spectral state, we analyzed
the data from the Proportional Counter Array instrument (PCA), which is
sensitive in the range 2--60~keV. 
To calculate the CD of Sco~X-1 we extracted
soft color and hard color lightcurves in 64\,s bins, where the soft color is
defined as the ratio between the count rate in the energy bands
[4.08--6.18~keV]/[1.94--4.05~keV] and the hard color is the ratio
[8.32--16.26~keV]/[6.18--8.32~keV]. These energy ranges are equivalent to
those used in previous studies of the source \citep[e.g.,][]{2007ApJ...667..411D}.
The resulting CD for the whole observational campaign is shown in
Figure~\ref{fig:ztrack}, where it can be seen that the source practically
covered the full double banana-shaped track during the observations.
We selected the top part of the upper banana of the CD
(indicated by a gray box in Figure~\ref{fig:ztrack}) as corresponding to the HB.
This selection is supported by previous observations of Sco~X-1 with \emph{RXTE},
where periods with similar CD selections were found to show the hard X-ray
power-law component characteristic of the HB state
\citep{2006ApJ...649L..91D,2007ApJ...667..411D}.

\section{Results}

A night-by-night analysis of the X-ray spectral states of Sco~X-1 showed that the
source did not move extensively along the Z track during any of the individual
observation nights, as can be seen in Figure~\ref{fig:ztrack}. During three of
the six VHE observations the source was in the HB (MJD 55332, 55334 and 55335),
as indicated in the observation log of Table~\ref{tab:log}. 

The total significance (following the definition of
\citealt{1983ApJ...272..317L}) of the gamma-ray signal coming from Sco~X-1 for
the complete data set of MAGIC observations, which amounts to 7.75~hr of
effective time, is $S=0.52\sigma$. For the complete dataset, the computed flux
upper limit is $2.4\times10^{-12}\,\cms$ above 300\,GeV. The simultaneous X-ray
observations allowed us to select the VHE data corresponding to each X-ray
state. We performed a signal search for those periods when the source was in the
HB, but no significant excess was found, with a flux upper limit of
$3.4\times10^{-12}\,\cms$ above 300\,GeV. A search for a signal
in the rest of the data set, when the source was in NB and FB states,
did not produce a positive result either. A summary of the X-ray-state
selected results for the VHE data is shown in Table~\ref{tab:uls}.

\begin{deluxetable}{ccccc}
    \small
\tablecaption{Integral Upper Limits for Selected X-ray States\label{tab:uls}}
\tablehead{
\colhead{X-ray State} & \colhead{Effective Time} & Significance & \multicolumn{2}{c}{UL $(>300\,\mathrm{GeV})^\mathrm{a}$}  \\
& \colhead{(hr)}   &  \colhead{$\sigma$}            & \colhead{(\cms)}   & \colhead{C.U.}
                      }
\startdata
All   & 7.75 & $0.52$ & $2.4\times10^{-12}$ & 1.9\%\\
HB    & 3.58 & $0.72$ & $3.4\times10^{-12}$ & 2.7\%\\
NB/FB & 4.17 & $0.08$ & $2.8\times10^{-12}$ & 2.3\%
\enddata
\tablenotetext{a}{Upper limits at 95\% confidence level.\vspace{1em}}
\end{deluxetable}

Additionally, a night-by-night analysis was performed. In none of the six
observations {was a significant signal found; the corresponding flux upper
limits are between} $2.0\times10^{-12}\,\cms$ and $7.9\times10^{-12}\,\cms$, as shown
in Table~\ref{tab:log}.

\begin{figure}
    \centering
    \includegraphics{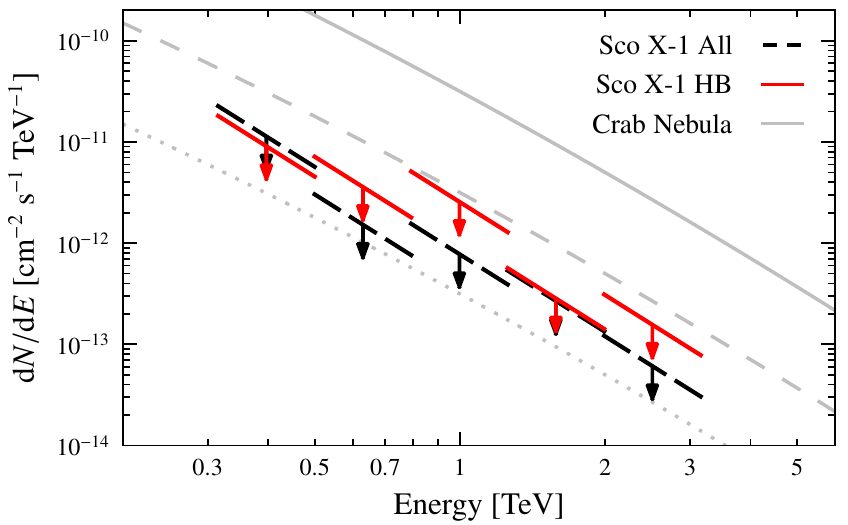}
    \caption{Differential upper limits for all the observations (black, dashed)
    and for the days in the HB (red). The slope of the bar indicates the assumed
    power-law photon index in the calculation of the upper limit. The Crab
    Nebula spectrum \citep{2008ApJ...674.1037A} is shown for comparison, as well
    as its 10\% (dashed) and 1\% (dotted) fractions.}
    \label{fig:sed}
\end{figure}

In Figure~\ref{fig:sed}, we show the differential flux upper limits computed for
the HB state and the whole data set to serve as a constraint for future
theoretical modeling of the VHE emission of the source. {Differential
photon flux upper limits may also be used to obtain energy flux upper limits.
Taking $\Gamma=3$, we obtain integral energy flux upper limits of $3.3$, $5.6$,
and $5.4\times10^{-12}\,\ergcms$ above 300\,GeV for the whole data set, the HB
data and the NB/FB data, respectively.}

\section{Discussion}\label{sec:disc}

To put in context the MAGIC results and understand the potentialities of Sco~X-1
to produce VHE emission, in this section we discuss some of the possible
emission scenarios. We consider the energy budget available from the
relativistic jet and the processes that could give rise to VHE emission either
at its base or further away from the compact object. Finally, we compare
Sco~X-1 with other microquasars and the detected gamma-ray binaries, and conclude
with a brief summary.

The emitter properties will be mostly constrained by the jet power, which
dictates the maximum energy budget available, and the emitter size and location.
The jet luminosity ($L_\mathrm{j}$) can be fixed assuming that the power-law
hard X-rays come from the jet, and assuming an X-ray luminosity to total jet power
ratio of 0.01--0.1. Given that the power-law luminosity is
$L_\mathrm{ntX}\sim 10^{35}\mbox{--}10^{36}$~erg~s$^{-1}$, a value of $L_\mathrm{j}\sim
10^{37}$~erg~s$^{-1}$ seems reasonable, and is also high enough to account for the
radio emission from the source \citep{2001ApJ...558..283F}.
The VHE luminosity upper limits in the HB show that Sco~X-1 has a maximum VHE
luminosity to jet power ratio of $L_\mathrm{VHE}/L_\mathrm{j}\lesssim 10^{-3}$.
This value is slightly below the ratio inferred for the Cygnus~X-1 flare
\citep{2007ApJ...665L..51A}, and similar to the upper limit of Cygnus~X-3
\citep{2010ApJ...721..843A}, {but an order of magnitude above the upper limit for
GRS~1915+105 \citep{2009A&A...508.1135H}.}
However, given the transient nature of these
sources, the orbital coverage and total length of the observational campaign
play a role in their eventual detection. We note that the campaign presented
here is relatively short and does not provide a complete orbital coverage so it
does not rule out the possibility of VHE flares from Sco~X-1.

Assuming that the jet of Sco~X-1 can indeed accelerate particles efficiently,
there are several possible reasons to explain the MAGIC non-detection. We will
begin by considering the possibility that particles are accelerated in the jet
base and the power-law hard X-rays are emitted there. At this location, it may
be difficult to accelerate particles beyond 100\,GeV because of strong radiative
cooling. The emitted spectrum is likely to be very soft because of the dominant
synchrotron cooling and IC occurring deep in the Klein-Nishina regime.
In addition, the radiation from 10\,GeV to 1\,TeV would likely be absorbed
through pair creation by the intense radiation field from the accretion disc. At
this location, even lower energy GeV photons would be absorbed. However, if the
magnetic field is low enough and electromagnetic cascading
\citep[e.g.,][]{1985Ap&SS.115...31A} is efficient, absorption at the jet base
owing to the disc photon field might not be so relevant. Another
source of absorption is the stellar photon field, enhanced through accretion
X-ray irradiance of the stellar surface, and could be optically thick for VHE
gamma-rays in some orbital phases. Note however that MAGIC has not detected the
source even for those phases in which this source of gamma-ray absorption should
be negligible: on MJD~55334 the source was in the HB
and the orbital phase range of the observation was $\phi$=0.03--0.12, for which
the expected gamma-ray opacity is $\tau\sim0.1$ \citep{2010A&A...514A..61B}. 
The combined effect of a steep gamma-ray spectrum and pair creation absorption
would diminish the chances of a VHE detection, particularly for an observation
at high zenith angles and correspondingly high energy threshold as presented
here.


On the other hand, radiation at VHE produced farther from the compact
object, at distances above $10^8$\,cm, could be eventually detected. The SSC
channel, efficient for intermediate values of the magnetic field, would yield a
less steep spectrum and may occur in regions where the emitter and the
environment are optically thin to VHE photons. IC with the X-ray enhanced
stellar photon field could also yield significant and less steep VHE radiation
and would be most efficient for optical depths of order of unity. According to
the calculations of \cite{2010A&A...514A..61B}, this would take place around
orbital phases close to 0.3 and 0.7, but we did not detect such emission on
MJD~55335 at orbital phases 0.25--0.39.
At these high altitudes in the jet, GeV photons could easily escape the system
and produce detectable high-energy gamma-ray emission. The eventual detection of
Sco~X-1 by \emph{Fermi} or \emph{AGILE}, plus a non-detection with low-enough
upper limits at VHE, would likely favor the scenario where gamma-ray radiation
is emitted from a location far away from the neutron star but with a spectrum 
too steep for a VHE detection. Synchrotron cooling dominance could explain the
steep spectrum and would mean that the power-law hard X-rays have a synchrotron
origin. {We note that the energy flux of an SSC/IC spectral component
peaking at 1\,GeV, with $\Gamma=3$ at higher energies and consistent with our
upper-limit during the HB, would only represent, at most, 0.8\% of the Eddington
luminosity for the neutron star.}

For leptonic emission scenarios, such as the SSC and external IC described
above, it can be assumed that X-ray emission has a synchrotron origin and VHE
gamma-rays have an IC origin. The VHE to non-thermal X-ray luminosity ratio is
then a sensitive indicator of the importance of synchrotron cooling and
ultimately of the magnetic field of the emitter. For the case of Sco~X-1, and
considering only the hard X-ray power law as non-thermal X-ray emission, this
ratio has a maximum value of $L_\mathrm{VHE}/L_\mathrm{ntX}\lesssim 0.02$ for
the HB. On the hitherto detected gamma-ray loud X-ray binaries this ratio is
much higher. For the case of LS\,I\,+61\,303, for example, the VHE to X-ray
ratio is between 0.5 and 1 during the X-ray and VHE peak at phases
$\phi=$0.6--0.7 \citep{2009ApJ...706L..27A}. Even considering the average fluxes
for the full simultaneous campaign during 60\% of an orbit the ratio is 0.26.
In both cases, the VHE to X-ray luminosity ratio is higher than the value we
obtained for Sco~X-1 by more than an order of magnitude. 
However, as mentioned above for the comparison with other microquasars, the
short total length of the observational campaign presented here prevents us from
making absolute comparisons because of the possible transient nature of the
source.
On the other hand, if there is no such flaring behavior, the difference in the
VHE to X-ray luminosity ratios indicates that 
the underlying physics is significantly different than for LS\,I\,+61\,303 or
LS\,5039.

In conclusion, our results place the first upper limits on the VHE gamma-ray
emission from Sco~X-1 in all of the X-ray states of the source. If Sco~X-1 is
indeed capable of producing TeV emission, either a longer observational
campaign, with better orbital coverage (to probe phase-dependent effects such as
absorption, cascading, etc.), or an instrument with a significant increase in
sensitivity, such as the future {Cherenkov Telescope Array (CTA)}, might
be required to detect it. Our upper limits also indicate that the underlying
VHE emission physics may be inherently different in the case of Sco~X-1 and the
detected gamma-ray binaries.

\acknowledgments
This research project has made use of data collected by NASA's
\dataset[ADS/Sa.RXTE#P/95341-01]{\emph{RXTE}}. We thank the \emph{RXTE}
scheduling team for their help in coordinating the simultaneous observations.
We also thank the Instituto de Astrof\'{\i}sica de Canarias for
the excellent working conditions at the Observatorio del Roque de los Muchachos
in La Palma. The support of the German BMBF and MPG, the Italian INFN, the
Swiss National Fund SNF, and the Spanish MICINN is gratefully acknowledged. This
work was also supported by the Marie Curie program, by the CPAN CSD2007-00042
and MultiDark CSD2009-00064 projects of the Spanish Consolider-Ingenio 2010
programme, by grant DO02-353 of the Bulgarian NSF, by grant 127740 of the
Academy of Finland, by the YIP of the Helmholtz Gemeinschaft, by the DFG Cluster
of Excellence ``Origin and Structure of the Universe'', and by the Polish MNiSzW
grant 745/N-HESS-MAGIC/2010/0.

\emph{Facilities:} \facility{RXTE (PCA)}, \facility{MAGIC}

\bibliographystyle{apj}
\bibliography{scox1-ads}

\end{document}